\documentclass[aps,prl,preprint,superscriptaddress]{revtex4-1}

\usepackage{graphicx}
\usepackage[version=3]{mhchem} 
\usepackage[T1]{fontenc}     
\usepackage [latin1]{inputenc}  
\usepackage{color}
\usepackage{xcolor}

\newcommand{\ie}{\textit{i.e.}}
\newcommand{\eg}{\textit{e.g.}}
\newcommand{\fig}[1]{FIG.~\ref{#1}}

\newcommand{\etal}{\textit{et al.}}

\newcommand{\tsup}[1]{\textsuperscript{#1}}
\newcommand{\tsub}[1]{\textsubscript{#1}}

\begin{document}


\title{Gone in 23 Attoseconds: Charge Transfer in Resonantly Core Excited Black Phosphorous}



\author{Fredrik O. L. Johansson}
\email[]{fredrik.johansson@physics.uu.se}
\affiliation{Division of Molecular and Condensed Matter Physics, Department of Physics and Astronomy, Uppsala University, Box 516, SE-751 20 Uppsala, Sweden}

\author{Yasmine Sassa}
\affiliation{Department of Physics, Chalmers University of Technology, SE-412 96 G\"{o}teborg, Sweden}

\author{Tomas Edvinsson}
\affiliation{Department of Engineering Sciences - Solid State Physics, Uppsala University, Box 534, SE-751 21 Uppsala, Sweden}

\author{Andreas Lindblad}
\affiliation{Division of Molecular and Condensed Matter Physics, Department of Physics and Astronomy, Uppsala University, Box 516, SE-751 20 Uppsala, Sweden}


\date{\today}

\begin{abstract}
How fast processes can we measure? Attosecond physics address the limit of measurable time in science. Atomic X-ray excited
states offers a way to study extremely fast dynamics with chemical specificity.
In black phosphorous an X-ray excited electron can relocate in 22.7 attoseconds. Using the lifetime of the P \textit{1s} core-hole as time-base, the radiationless decay spectrum can be used to study charge transfer processes on the time-scale of the atomic unit of time (24 attoseconds). We demonstrate that the technique can be extended to within a few percent of the core hole's lifetime, an order of magnitude smaller than previously thought.
%
\end{abstract}



\maketitle


\begin{figure}[]
\centering
\includegraphics[]{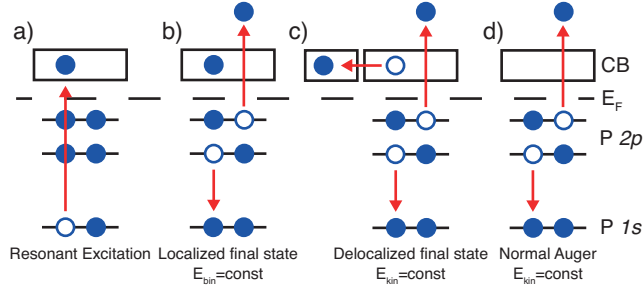}
\caption{Schematic representation of resonant excitation and subsequent possible auto-ionization processes. Filled green circles represents electrons in an occupied level, hollow circles holes. In c) the second box represents the conduction band on another atom.}
\label{fig:1}
\end{figure}

An electronic system to respond to an electromagnetic pulse in the 10 to 100 attoseconds \cite{hassan2016optical}; a regime out of range for studies of electronic processes. Experimental methods to go below 100 attoseconds are needed to study fundamental processes in matter-radiation interactions  -- for instance, rapid electron dynamics in semiconductor systems, \eg~how excited electrons cause a short time-scale electronic response and how this affect other electronic processes and lattice response on comparably longer time-scale (60 femtoseconds) \cite{schultze2014attosecond}.

Ultra-fast processes are also of profound importance for the characterization of electron dynamics in functional materials, \eg{} photovoltaics \cite{FJ_JPCC}. These processes, occurring on the attoseconds timescales \cite{RevModPhys.81.163}, are commonly investigated using laser pump-probe spectroscopies where processes are tracked as a function of the temporal change of the delay between the pump and probe \cite{cappel2016electronic}. To resolve these processes the laser pulse, the delay and the probe together must hit the sample within the same time-frame as the duration of the dynamic process. This then sets a lower limit of this technique slightly above 100 attoseconds but with limitations on the pulses intensities \cite{0034-4885-80-5-054401}. To investigate even faster processes, some other technique must be used. It has been shown that for direct photoemission a delay of 21 attoseconds can be observed comparing non-resonant electron emission from the \textit{2s} and \textit{2p} orbitals of neon -- with the system excited by the same photon pulse \cite{Schultze1658}. From a W(110) solid the delay between core level \textit{4f} electron emission and emission from electrons at the Fermi level is about 90 attoseconds \cite{cavalieri2007attosecond}.

\begin{figure*}[]
\centering
\includegraphics[width=16cm]{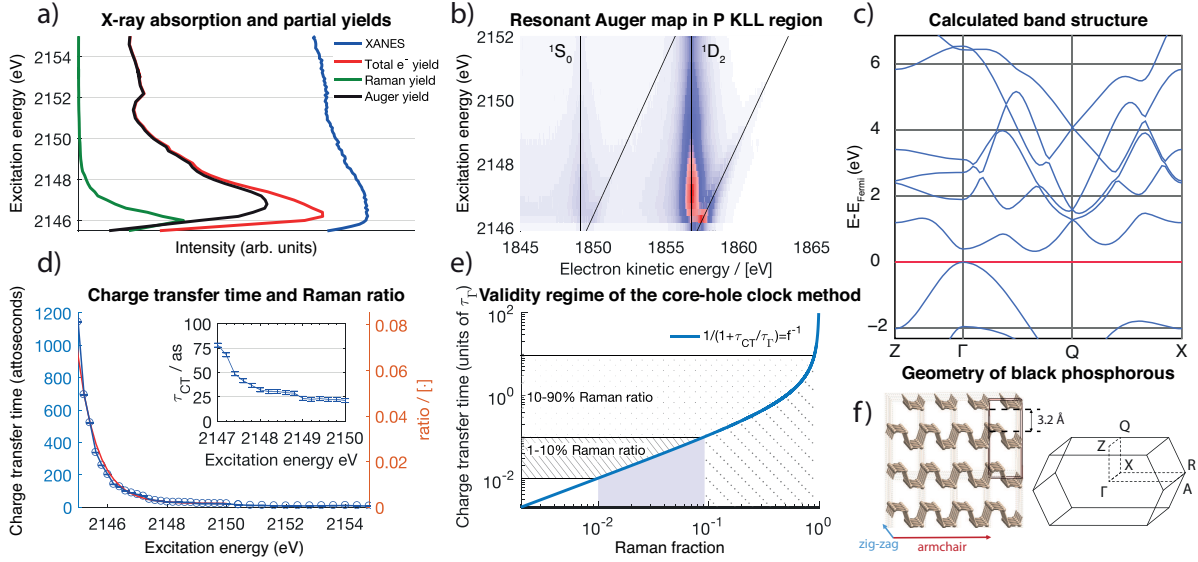}
\caption{\label{fig:2}  a) shows the XAS and partial electron yields for the auto-ionization processes. b) shows the resonant auger map over the P $1s$ to $3p_z$ resonance 
sloped lines highlight the Raman feature and vertical lines the charge transfer feature
for both the \tsup{1}S\tsub{0} and the \tsup{1}D\tsub{2} Auger channels. c) shows the calculated states of the unoccupied bands. d) shows the calculated charge transfer times, with an exponential fit to the data, note the 2 y-axises with charge transfer times on the left and Raman ratio at the right. The inset shows the
the region where the charge transfer times are shortest. e) shows a plot of the electron transfer time in units of the core-hole lifetime as function of the Raman fraction in the decay spectra. The dashed grey area indicates the previously thought limits of the core-hole clock method from 0.1 to 0.9 Raman fraction as described by Wurth and Menzel in 2000\cite{Menzel2000}, the 1-10\% area indicate our new suggested lowering of the core-hole clock limit. f) depicts the crystal structure and the unit cell in reciprocal space for bulk black phosphorous.}
\end{figure*}

Here we use core-hole clock spectroscopy, using the life-time of an X-ray excited state as an internal time reference for dynamic processes studied via the electron emission produced in the decay of that state \cite{bruhwiler2002charge}. F\"ohlisch and co-workers demonstrated that a charge transfer time of 320 attoseconds can be determined for sulfur atoms on a Ru(0001) surface \cite{fohlisch2005direct}. Further applications of the technique have identified attosecond electron charge transfer dynamics on the 100's of attosecond time-scale in liquid and frozen water \cite{PhysRevLett.99.217406}, the DNA backbone \cite{PhysRevLett.99.228102}, polymers for organic photovoltaics \cite{doi:10.1021/jp508010u,FJ_JPCC} and phosphorous doped graphene \cite{C5RA12799H}. 

Single crystal Black Phosphorous (P) -- a material of which monolayers have shown promise as novel 2D semiconductor devices \cite{gusmao2017black} -- is here used as a model material to investigate ultra fast excitation and charge transfer.
In this report we have used tender X-rays from a synchrotron source to resonantly prepare a core-excited system, this meta-stable state may decay by the emission of an electron. 
The spectrum of the emitted electrons contain information on how fast competing processes need to be to occur during the life-time of the core-hole. We demonstrate a charge transfer time of 23 attoseconds, a time shorter than the atomic unit of time, 24.19 attoseconds \cite{Tiesinga2020}. This is an unprecedented charge transfer time an order of magnitude faster than in any other previously studied materials system and also pushes the validity range of the core-hole clock method down to percent of the core-hole lifetime. 

The experiment was conducted at the KMC-1 beamline and HIKE end-station at Helmholtz-Zentrum Berlin, BESSY II \cite{Schaefers2007,Gorgoi2009}. 
The light is linear horizontal polarized with the polarization in the plane of the samples surface normal. All measurement were done in normal emission towards the respective analyzer and in resonant Raman conditions -- the photon bandwidth being smaller than the lifetime of the core-excited state \cite{piancastelli2014core}.
The Black phosphorous single crystal is the same as characterized in reference \cite{DAVID2019144385}.

Hybrid density functional theory was utilized with and without addition of van der Walls interaction for Black phosphorous. The systems were fully geometry optimized to the electronic ground state using a 12x12x12 grid using the Monkhorst-Pack approach for both the geometry optimizations and property calculations. The electronic effect upon a Z+1 substitution defect was calculated using a 2x2x2 super cell containing 128 atoms. The basis set for P and S (Z+1 substitution atom) was based on a triple-zeta accuracy with polarization quality \cite{peintinger2013consistent}. All calculations were performed using the the Crystal17 code \cite{CRYSTAL,Dovesi2017}


A resonant core excitation from the P \textit{1s} orbital to the unoccupied bands built up from P \textit{3p} orbitals create a meta-stable core-excited state (\fig{fig:1}a) that decay by autoionization (\fig{fig:1}b-c). A non-resonant excitation with an X-ray energy above the ionization potential of the system create a core ionized state which may decay by emission of a valence electron bringing the system into an Auger like final state (\fig{fig:1}d) \cite{Meitner1922,auger1923rayons}. The energy of the Auger electron depend only on the energy of the orbitals involved, not on the X-ray energy. This is a non-resonant process with constant kinetic energy in the electrons' spectra. 
Following a resonant core-excitation two processes compete, the Raman-like decay involving the core-excited electron itself (\fig{fig:1}b) and the delocalized Auger-like charge-transfer decay (\fig{fig:1}c). These features are separated in kinetic energy in the Auger decay region as seen in \fig{fig:2}b. The chosen Auger electron kinetic energy regions is P KL\tsub{2,3}L\tsub{2,3} (\tsup{1}D\tsub{2}). The Raman-like localized spectator decay, is a coherent process in which the emitted electron has a kinetic energy that is proportional of the incoming photon energy (from here on called the Raman feature, notation following reference \cite{FOHLISCH2003107}) and has a final state of 2h1e (2 holes, 1 electron, $2p^{-2}3p^{+1}$). The Auger-like decay is incoherent and has the same final state as the normal (non-resonant) Auger decay, the energy of the emitted electron has a constant kinetic energy and a final state of 2h ($2p^{-2}$).
The ratio between the localized Raman- and the de-localized charge transfer features ($I_{Raman}/I_{CT}$) is called the Raman ratio and is then used to calculate the charge transfer time ($\tau_{CT}$) from $\tau_{CT}=I_{Raman}/I_{CT}\cdot\tau_{1s}$ where the Raman-ratio is multiplied with the lifetime of the P \textit{1s} core-hole, $\tau_{1s}$.







\fig{fig:2}a displays the recorded X-ray absorption spectrum of black Phosphorous around the P K-edge resonance located at approximately 2146 eV, our fluorescence yield data agreeing with those of Hayasi \etal{} \cite{hayasi1984electronic}. This is used to determine the energy position of the P $1s \to 3p_z$ resonance. The excited electron will predominantly populate the $3p_z$, since the X-rays from the synchrotron light source is linearly polarized in the same plane as the sample surface normal (see \eg~reference \cite{FOHLISCH2003107}).

\begin{figure}[tb!]
\centering
\includegraphics[]{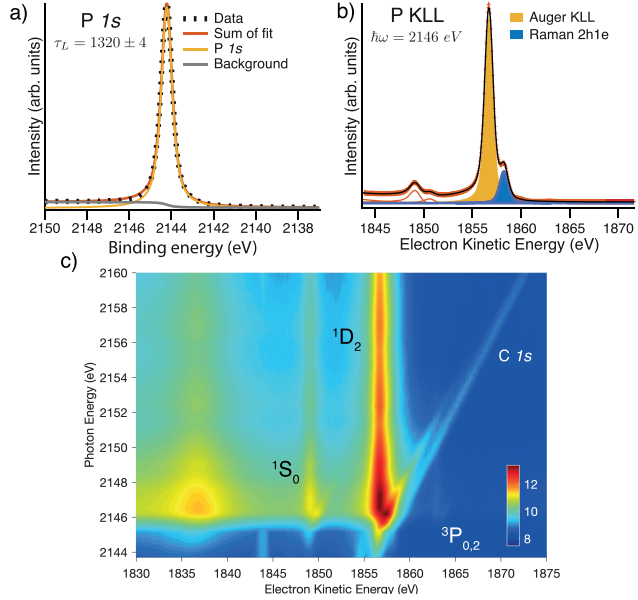}
\caption{\label{fig:xps}  a) shows the fit of the P \textit{1s} photoelectron spectrum. In panel b) the Auger spectrum recorded at 2146 eV photon energy, the dispersive and non-dispersive features in orange and blue respectively. Panel c) shows the full resonant Auger map with the rows interpolated and the intensity on a logarithmic scale to highlight the details with lower intensities.}
\end{figure}

A set of Auger spectra
were measured starting from an incident photon energy of 2143 eV to follow their evolution throughout the resonance, the spectrum recorded at 2146 eV photon energy can be seen in \fig{fig:xps}b. These are stacked together to form the resonant Auger map seen in \fig{fig:2}b and \fig{fig:xps}c. The resonant Auger map reveals several components: two Auger channels, the \tsup{1}D\tsub{2} (2h) KL\tsub{2}L\tsub{3} at 1857 eV kinetic energy and the \tsup{1}S\tsub{0} (2h) KL\tsub{1}L\tsub{1} at 1850 eV kinetic energy. The faint \tsup{3}P\tsub{0,2} (2h) KL\tsub{3}L\tsub{3} at 1865 eV kinetic energy can be seen in \fig{fig:xps}c where the intensity scale is logarithmic and the spectra interpolated to highlight these features.
The two Auger channels in \ref{fig:2}b are marked with vertical lines. The dispersive spectator states (2h1e) are marked with tilted lines that follows linearly with the incident photon energy. 
As can be seen in \fig{fig:xps}c, starting at 1856 eV kinetic energy (at 2143 eV photon energy), another dispersive feature can be seen -- the C \textit{1s} photoelectron line.  This is clearly distinguishable and separated from the spectator line and will not effect the determination of the charge transfer time.
The partial electron yields in the Raman and Auger channels belonging to the \tsup{1}D\tsub{2} region are included in \fig{fig:2}a.

To each measured Auger spectra a least squares fit with components for the \tsup{1}D\tsub{2} and the Raman feature have been performed. The charge transfer peak were first fitted from a normal Auger spectra, 
and this peaks position is then used when fitting all other spectra. The Auger peak on the resonance is asymmetric and is modeled with a Doniach-\v{S}uni\'{c} lineshape \cite{doniach1970many}. 
An example of a fitted spectra after the resonance can be seen in \fig{fig:xps}b with separated Raman and charge transfer components. 
The charge transfer time for each point on the resonance are presented in \fig{fig:2}d. The core-hole lifetime of the P \textit{1s} 
from the Lorentzian life-time obtained from the fit of the P \textit{1s} core level spectra (\fig{fig:xps}a)
measured at 6015 eV photon energy. The spectra is deconvoluted with a Shirley background \cite{shirley1972high} and Voigt lineshape for the photoline. The Gaussian broadening was determined by fitting a Au \textit{4f} spectrum, recorded with the same beamline settings as the P \textit{1s}. 
The Lorentzian broadening of Au $4f_{7/2}$ and $4f_{5/2}$ is  0.30 and 0.28 eV respectively \cite{PATANEN201159}. The Au \textit{4f} lines are fitted with these values fixed
 as Lorentzian component which gives an Gaussian broadening of 0.274 eV. The error in the experimental broadening was determined to be $\pm$2.7 meV by using the built-in Monte-Carlo error-fit function of the SPANCF IGOR package. The P \textit{1s} spectrum was then fitted this Gaussian broadening resulting in a Lorentzian broadening of 0.499 eV. The core-hole lifetime ($\tau$) is calculated to 1320 $\pm$4 attoseconds using the Heisenberg uncertainty principle, $\tau=\hbar/\Gamma$.
At the peak resonance energy the charge transfer time is on the same order as the core-hole lifetime. When the excitation energy deviate from that, the charge transfer time exponentially decreases and reaches a steady state of 22.7 ($\pm$4) attoseconds.
This truly ultra fast charge transfer time is shorter than the atomic unit of time (1 a.u. of time is 24 attoseconds). 
A resonantly populated meta-stable state decays exponentially if we assume a weak coupling to the rest of the system \cite{sanchez2007first}. 
This is also consistent with the behavior of transmission of an electron tunneling through a triangular barrier \cite{forbes2011transmission}.  



\begin{figure}[htb!]
\centering
\includegraphics[]{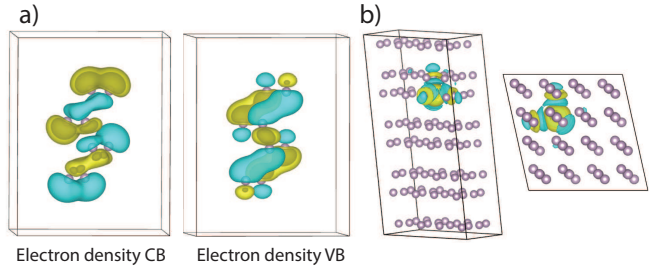}
\caption{\label{fig:3} a) The calculated conduction band and valence band electron densities. b) Differential plot of the electron density in Black Phosphorous when one phosphor atom i substituted to a Z+1 defect (\ie ~a sulfur atom).}
\end{figure}

A resonant excitation from the P $1s$ to the $3p_z$ orbital can be conceptualized as either a population of a virtual state where the core-hole potential is assumed to be instantaneously screened at the virtual level or a creation of a Z+1 defect at the core-excited site. The dynamics of the electron is then determined by the electronic structure (\fig{fig:3}b) in the new instantaneous situation where atomic nuclei are contributing but are effectively frozen at these fast time scales. The structure of black phosphorous is shown in \fig{fig:2}f, where the direction of the zig-zag and armchair directions is shown as well as the inter-layer spacing. The dynamics of the electrons after population of the virtual states, with assumption of screened core hole potential, would then be instantaneously affected by the new electronic situation. Since black phosphorous is highly anisotropic, naturally across the layers, but also in the zig-zag and arm-chair direction, the energy landscape will be markedly different in the different directions as well as depend on at which energy it is excited to. 

 The band structure in \fig{fig:2}c and the electron density in~\fig{fig:3} have been calculated using density functional theory. With a high excitation cross-section at the $\Gamma$-point and considering the ultra-fast time scales, the crystal momentum change can be neglected ($\Delta$k=0) and processes located vertically at the $\Gamma$-point (k=0) are considered in an effective mass analysis of the excited electron. 
The effective mass of the excited electron can be determined from the second derivative of the electron dispersion of the virtual orbitals around the symmetry point considered.



With increasing excitation energy the curvature of the bands near the $\Gamma$-point become smaller, which reflect a heavier effective mass. Higher mass implies slower dynamics, typically transmission for quantum mechanical tunnelling depend on the wave-number which is proportional to the square root of the mass \cite{landaulifshitz} -- this is also manifested in proton transfer, where dynamics are slower for deuterium \cite{C2CP23615J}.

 
 Considering the effect of the core hole potential, a Z+1 approximation is utilized, which for phosphor is analogous to a sulfur defect \cite{guo2015vacancy}, where a substitutional Z+1 defect in black phosphorous puts the extra electron in the \textit{p\tsub{z}} orbital \cite{li2015structures}. Another way of adding electrons to the system is alkali metal doping \cite{kim2015observation}. Kim and co-workers have shown that a defect can create an electric field, which separates the valence band and conduction band spatially. The resulting electron density is depicted in \fig{fig:3}a where the spatial separation that put the valence- and conduction band on separate layers, again consistent with the single exponential decay of the resonant population which we observe. From the theory, as both a virtual orbital analysis and an approach considering the core-hole potential via a Z+1 approximation, the instantaneous electronic effect of the core-hole effective potential show altered energy landscape within a layer of black phosphorous before subsequent lattice relaxation.


Charge transfer or delocalisation times ($\tau_{CT}$) studied with the core-hole-clock method are often said to be accessible in a range between one tenth and ten times the core hole lifetime ($\tau$). These limits have arisen from where the Raman fraction, $f={I_{Raman}}/[{I_{Raman}+I_{CT}}]$, is considered to be experimentally determinable (dotted in \fig{fig:2}e) -- as stated by Wurth and Menzel \cite{Menzel2000}. 
Plotting the Raman fraction on a log-log scale we indicate, in grey in \fig{fig:2}d, the range down to 1 \% of the core-hole lifetime. The per-cent range is well above the detection limit for electron spectroscopies \cite{shard2014detection}. In fact, the fastest charge transfer time determined here corresponds to a Raman fraction of $1.7\pm0.2\%$. We can access this range since the Raman and KL\tsub{2}L\tsub{3} Auger features are well separated in our spectra, see \fig{fig:xps}b. 
A vital condition for measuring ultra-short processes by this approach is thus to choose a system with well separated Raman and Auger features.
In the case of Black phosphorous the total uncertainty in the time-determination (uncertainty of the core hole life-time combined with that of the fraction) is about 4 as.
The extension of this range here demonstrate that we may study processes on a time-scale one magnitude shorter than previously thought.


The timescale in our result is in the 10\% range of previously fastest charge transfer times, the timescale has been touched before, as can be shown in the work of Wang \etal{} \cite{PhysRevA.50.1359} whom recorded the resonant Auger spectrum in the P KLL kinetic energy region over the phosphor K-edge resonance with fitting of the resonant (dispersing) and normal Auger components. Using their results and with a core-hole life time of 1830 attoseconds, we find charge transfer times that exhibit an exponential decay after the resonance. The exponential fit to the data reaches a steady state of 29.7 attoseconds after the resonance albeit with a substantial error-bar. Using the data from reference \cite{BABA1994896} measured in the Si KLL region in the vicinity of the Si \textit{1s} resonance of \ce{SiO2} we can, using the tabulated core hole lifetime of Krause \etal{} \cite{krause1979natural} ($\Gamma_{\ce{Si} \textit{1s}}=0.48 \ce{eV}$) deduce a charge transfer time of 50 attoseconds, with large errorbars just as in the case of \ce{InP}. 





In conclusion, we have demonstrated that we can access truly ultra-fast processes down to per cent levels of the core-hole lifetime. Using the P KLL Auger spectra allows us to extend the range of available Raman fractions that we can study by an order of magnitude --- thus accessing time scales of truly ultra-fast processes close to the atomic time unit (1 a.u. of time is 24 attoseconds). For black phosphorous the charge delocalization times is determined to 22.7$\pm$4 attoseconds, we expect charge transfer times for row 3 elements measured by KLL core-hole clock spectroscopy to exhibit the same contrast between the normal Auger and resonant Raman channels. All row 3 elements are interesting for wide application for opto-electronic devices. Core-hole clock spectroscopy can thus contribute to fundamental understanding of electronic properties of semi-conductor devices and junctions built up from those materials.

\begin{acknowledgments}
A.L. acknowledges the support from the Swedish Research Council (grants no. 2014-6463 and 2018-05336) and Marie Sklodowska Curie Actions (Cofund, Project INCA 600398). Y.S. is fully supported by a VR neutron project grant (BIFROST, Dnr. 2016-06955) as well as a VR starting grant (Dnr. 2017-05078). F.J. acknowledges financial support from the K G Westman foundation. We thank HZB for the allocation of synchrotron radiation beamtime and Roberto Felix Duarte for assistance at the KMC-1 beamline.

\end{acknowledgments}


\providecommand{\noopsort}[1]{}\providecommand{\singleletter}[1]{#1}%

\end{document}